\documentclass[10pt,letterpaper]{article}
\usepackage{opex3}
\usepackage{cite}
\usepackage{color}
\usepackage{epsfig,dsfont,amssymb,amsmath,amsthm,amsfonts,amsbsy,mathrsfs}
\usepackage{graphicx}
\usepackage{amsmath}
\usepackage{amssymb}
\usepackage{lipsum}
\usepackage{amsfonts}%
\begin{document}

%%%%%%%%%%%%%%%%%% title page information %%%%%%%%%%%%%%%%%%
\title{Optomechanical entanglement under pulse drive}

\author{Qing Lin$^{1,3}$ and Bing He$^{2,4}$}

\address{$^1$Fujian Provincial Key Laboratory of Light Propagation and Transformation, College of Information Science and Engineering, Huaqiao University,
Xiamen 361021, China\\
$^2$Department of Physics, University of Arkansas, Fayetteville, AR 72701, USA\\
$^3$qlin@hqu.edu.cn\\
$^4$binghe@uark.edu}



\begin{abstract}
We report a study of optomechanical entanglement under the drive of one or a series of laser pulses with arbitrary detuning and different pulse shapes. Because of the non-existence of system steady state under pulsed driving field, we adopt a different approach 
from the standard treatment to optomechanical entanglement. The situation of the entanglement evolution in high temperature is also discussed.
\end{abstract}

 % REPLACE WITH CORRECT OCIS CODES FOR YOUR ARTICLE, MINIMUM OF TWO; Avoid using the OCIS codes for “General” or “General science” whenever possible.

%%%%%%%%%%%%%%%%%%%%%%% References %%%%%%%%%%%%%%%%%%%%%%%%%

%%%%%%%%%%%%%%%%%%%%%%%%%%  body  %%%%%%%%%%%%%%%%%%%%%%%%%%
\section{Introduction}
Due to their various interesting properties, optomechanical systems (OMS) are under extensive researches over recent years \cite{review1,review2,review3,review4,review5}. They are regarded as good platforms for realizing high precision detection \cite{pm1,pm2}, e.g. detection of gravitational wave \cite{pm3,pm4}, ultra-low temperature cooling, e.g. cooling the nano-mechanical oscillator to its ground-state \cite{cooling-1,cooling-2,cooling-3,cooling-4}, and macroscopic quantum states, e.g. ``cat state" or macroscopic quantum entanglement \cite{ET-1,ET-2,ET-3,ET-4,ET-5,ET-6,ET-7,ET-8,ET-9}, and others.
An OMS is nonlinear by nature, so it is difficult to find its exact dynamical evolution in the regime of strong optomechanical coupling. In the weak coupling regime, the fluctuation expansion around the classical steady states is a standard method for 
studying any feature of an OMS, including optomechanical entanglement \cite{ET-1,ET-2}. With the replacements $\hat{a} \rightarrow \bar{a}+\delta \hat{a}$ and $\hat{b} \rightarrow \bar{b}+\delta \hat{b}$, the original cavity (mechanical) mode is expanded into the sum of the average value $\bar{a}$ ($\bar{b}$), as the steady state found from classical equations of motion, and its fluctuation $\delta \hat{a}$ ($\delta \hat{b}$). Then the initial Hamiltonian will be linearized so that the quantum Langevin equations about the fluctuations can be solved. The practical use of the fluctuation expansion approach, however, relies on the solutions of classical nonlinear dynamical equations, which often take the approximate forms. In the cases of continuous wave (CW) drives, such solutions 
are the classical steady states determined by Routh-Hurwitz criterion \cite{criterion}. 

The explicit situations when there is no steady state include the pulse and blue detuned CW driving fields. 
A possible way to deal with these situations is the evolution decomposition method \cite{He, LH}, which works in any regime without the restriction of the steady state condition. A number of new phenomena were discovered based on this approach. For instance, the entanglement under blue detuned drive is predicted to be higher and more robust against high temperature than that under red detuned drive. Moreover, it is found that quantum noise effect can be enhanced by drive intensity, and would destroy the entanglement accordingly, resulting in sudden death and sudden revival of OMS entanglement \cite{He,LH}.

Here we will apply this method to investigate the OMS entanglement created by pulsed drives. As it is known to all, a blue detuned drive would ``heat" an OMS. A blue detuned CW drive, with its longer interacting time with the system, will increase the system temperature continually. The decoherence accompanying the rising temperature would diminish and even destroy the entanglement from optomechanical coupling. If a pulsed drive is used instead, the heating of the system would not be accumulated to significant temperature, and thus have the corresponding decoherence safely neglected for the blue detuned drive. This is one motivation to consider optomechanical entanglement under pulsed drive.

OMSs under pulsed drives have been considered for quantum state preparation, cooling, quantum entanglement creation, as demonstrated by the recent experiments \cite{Sposition,pstate,cooling-p,Hofer,Vanner1,Machnes,Vanner2,HQY1,cats,HQY2}. In 2011, Hofer, et al. showed that the entanglement (measured by EPR variance) created by pulsed drive of blue detuned (with the detuning $\Delta_0=-\omega_m$) can be larger than the upper bound of entanglement due to red detuned CW drive \cite{Hofer}. Later, He, Kiesewetter, et al. demonstrated the existence of quantum steering in pulsed OMS \cite{HQY1} and also predicted 
that robust EPR steering (without requiring a low temperature reservoir) could be achieved by a square pulse \cite{HQY2}.

In this paper, we will present a detailed study of entanglement evolution under pulsed drives of different detunings and different shapes. With these evolutions, we will show that the created entanglement can even last for longer time than pulses themselves, and such entanglement is also robust against relatively high temperature when driven by blue detuned laser pulses. In addition, if a series 
of pulses continuously drive an OMS, nonzero entanglement could be preserved even when the gap time between the laser pulses is considerable.

\section{System Hamiltonian}

Here we consider an OMS driven by a single laser pulse or a series of laser pulses, which are given in the form $\sum_j E_j(t-j\cdot t_0)e^{i\omega_0 (t-j\cdot t_0)}$ with $j$ being the number of pulses, $t_0$ being the gap time between the pulses, and $\omega_0$ being the central frequency of the pulsed drive. The function $E(t)$ represents the profile of the pulse. Under the pulsed drive, the system Hamiltonian in the interaction picture with respect to $H_0=\omega_c \hat{a}^\dagger \hat{a}+\omega_m \hat{b}^\dagger \hat{b}$ $ (\hbar \equiv 1)$ is
\begin{eqnarray}
\vspace{-0.3cm} H_S(t)=-g\{ \hat{b} e^{-i\omega_m t}+ \hat{b}^\dagger e^{i\omega_m t}\}\hat{a}^\dagger \hat{a} +i\left(\sum_j E_j(t)\hat{a}^\dagger e^{i\Delta_0 (t-j\cdot t_0)}-\sum_j E^{*}_j(t)\hat{a}e^{-i\Delta_0 (t-j\cdot t_0)}\right)  \label{HS}
\end{eqnarray}
where $g$ is the optomechanical coupling factor, $\omega_c$ ($\omega_m$) is the cavity (mechanical) oscillation frequency, $\Delta_0=\omega_c-\omega_0$ is the detuning. In the above equation, the first two terms describe the interaction between mechanical mode and cavity mode, while the last terms represents the driving on the cavity.

The system we consider is an open system, i.e. the cavity and mechanical mode will damp with the rate $\kappa$ ($\gamma_m$). The coupling to the reservoir can be described as the stochastic Hamiltonian \cite{book}:
\begin{eqnarray}
H_D(t)=i\big(\sqrt{\kappa}\hat{a}^\dagger \hat{\xi}_c(t) +\sqrt{\gamma_m}%
\hat{b}^\dagger \hat{\xi}_m(t)\big )+H.c.  \label{HD}
\end{eqnarray}
where $\hat{\xi}_c$ ($\hat{\xi}_m$) is the stochastic Langevin noise operator. Under these factors, the joint evolution of the system and reservoir manifests in term of the evolution operator $U_S(t,0)=\mbox{T}e^{-i\int_0^t d\tau \big(H_S(\tau)+H_D(\tau)\big)}$. Since the coupling and dissipation processes in the evolution are non-commutative, it is impossible to get a closed form of this time order exponential. 

\section{Factorization of joint system-reservoir evolution operator}

Here we adopt a different approach from the standard fluctuation expansion method to study the system evolution. This approach is based on the factorization of evolution operator and is valid for arbitrary driving field including the pulsed ones \cite{He,LH}.
The factorizations for the evolution operator is given as follows \cite{He}:
\begin{eqnarray}
\mbox{T}e^{-i\int_{0}^{t}d\tau (H_{1}(\tau )+H_{2}(\tau ))} =\left\{\begin{matrix}
 &\mbox{T}e^{-i\int_{0}^{t}d\tau H_{1}(\tau )}~\mbox{T}e^{-i\int_{0}^{t}d\tau V_{1}^{\dagger }(\tau ,0)H_{2}(\tau )V_{1}(\tau ,0)}; \text{  Left Factorization} \\ 
 & \mbox{T}e^{-i\int_{0}^{t}d\tau V_{2}(t,\tau )H_{1}(\tau )V_{2}^{\dagger}(t,\tau )}~\mbox{T}e^{-i\int_{0}^{t}d\tau H_{2}(\tau )}; \text{  Right Factorization} 
\end{matrix}\right.
\end{eqnarray}
where $V_{k}(t,\tau )=\mbox{T}\exp \{-i\int_{\tau }^{t}d\tau ^{\prime
}H_{k}(\tau ^{\prime })\}$ for $k=1,2$. With these factorizations, the evolution operator 
can be simplified, so that it is possible to applied the factorized operators in succession to find the 
system observables that should be obtained by directly acting the original evolution operator $U_S(t,0)$ on system operators or system's initial quantum state. In addition to OMS, this method has been applied to other physical systems recently \cite{a1, a2, a3}.

The factorization procedure for our concerned system is as follows. At first, we implement a right factorization to separate the noise part $U_{D}(t,0)=\mbox {T}\exp \{-i\int_{0}^{t}d\tau H_{D}(\tau )\}$ out of the joint unitary operator:
\begin{eqnarray}
U_S(t,0)=\mbox{T}\exp\{-i\int_0^t d\tau U_D(t,\tau)H_S(\tau)U_D^\dagger(t,\tau)\}
~\mbox {T}\exp\{-i\int_{0}^t d\tau H_D(\tau)\}.
\label{one}
\end{eqnarray}
In the main Hamiltonian $U_D(t,\tau)H_S(\tau)U_D^\dagger(t,\tau)$, the cavity, mechanical mode operators $\hat{a}$, $\hat{b}$ 
are transformed to
\begin{eqnarray}
U_D(t,\tau)\hat{a}U_D^\dagger(t,\tau) =&e^{-\frac{\kappa}{2}(t-\tau)}\hat{a}+\hat{n}_c(t,\tau) &\equiv \hat{A}(t,\tau),  \label {Tra}\\
U_D(t,\tau)\hat{b}U_D^\dagger(t,\tau) =&e^{-\frac{\gamma_m}{2}(t-\tau)}\hat{b}+\hat{n}_m(t,\tau) &\equiv\hat{B}(t,\tau),\nonumber 
\end{eqnarray}
where $\hat{n}_c(t,\tau)=\sqrt{\kappa}\int_{\tau}^t d\tau'e^{-\kappa(\tau'-\tau)/2}\hat{\xi}_c(\tau')$ and $\hat{n}_m(t,\tau)=\sqrt{\gamma_m}\int_{\tau}^t d\tau'e^{-\kappa(\tau'-\tau)/2}\hat{\xi}_m(\tau')$ are the induced quantum noise operators satisfying the following commutation relations:
\begin{eqnarray}
\Gamma_c(t',\tau)=&[\hat{n}_c(t,t'),\hat{n}_c^{\dagger}(t,\tau)]&=e^{-\kappa(\tau-t')/2}-e^{-\kappa(t-\tau)/2}
e^{-\kappa(t-t')/2},\\
\Gamma_m(t',\tau)=&[\hat{n}_m(t,t'),\hat{n}_m^{\dagger}(t,\tau)]&=e^{-\gamma_m(\tau-t')/2}-e^{-\gamma_m(t-\tau)/2}
e^{-\gamma_m(t-t')/2}.\nonumber
\end{eqnarray}
The transformed system operators therefore satisfy the equal-time commutation relation $[\hat{A}(t,\tau), \hat{A}^{\dagger}(t,\tau)]=[\hat{B}(t,\tau), \hat{B}^{\dagger}(t,\tau)]=1$. 
After that, we factorize the drive term $U_E(\tau,0)=\mbox{T}\exp\{\int^\tau_0 dt^{\prime}\sum_j E_j(t^{\prime})\hat{A}^\dagger(t,t^{\prime}) e^{i\Delta_0 (t^{\prime}-j\cdot t_0)}\}$ to the left side, and the cavity mode in 
the rest operator is determined by the transformation
\begin{eqnarray}
U_E^\dagger(\tau,0)\hat{A}(t,\tau)U_E(\tau,0)&=&\hat{A}(t,\tau)+\sum_j \left[e^{-\frac{\kappa}{2}(t-\tau)}\int_0^\tau dt' E_j(t^{\prime})e^{i\Delta_0 (t^{\prime}-j\cdot t_0)}e^{-\frac{\kappa}{2}(t-t')} \right. \nonumber\\
&+& \left. \int_0^\tau dt' \Gamma_c(t',\tau)E_j(t^{\prime})e^{i\Delta_0 (t^{\prime}-j\cdot t_0)}\right]\nonumber\\
&\equiv & \hat{A}(t,\tau)+\sum_j D_j(\tau),
\label{displace}
\end{eqnarray}
Finally, we factorize the term $U_C(t,\tau)=\mbox{T}\exp\{ig\int^t_{\tau}dt^{\prime}\left[\hat{B}(t,\tau) e^{-i\omega_m t}+ \hat{B}^\dagger(t,\tau) e^{i\omega_m t}\right]\hat{A}^\dagger\hat{A}(t,t^{\prime})\}$ out of the optomechanical coupling operator from the right side. To the first order of the optomechanical coupling constant $g$, we obtain the following effective Hamiltonian,
\begin{eqnarray}
\tilde{H}_{OM}(\tau)&=&-g\left[\hat{B}(t,\tau) e^{-i\omega_m t}+ \hat{B}^\dagger(t,\tau) e^{i\omega_m t}\right]\nonumber\\
&\times & \left[\sum_j \hat{A}(t,\tau)D^{*}_j(\tau)+\sum_k \hat{A}^\dagger(t,\tau)D_k(\tau)+\sum_j\sum_k D^{*}_j(\tau)D_k(\tau)\right]. \label{eff}
\end{eqnarray}
which is a good approximation of the effective optomechanical coupling Hamiltonian for the weak coupling regime of an OMS.
The joint unitary operator is therefore factorized as
\begin{eqnarray}
U_S(t,0)=U_E(t,0)U_{OM}(t,0)U_C(t,0)U_D(t,0).
\end{eqnarray} 

Since the OMS we consider is an open system, the joint initial state is the following product state
\begin{eqnarray}
\chi(0)=\rho(0) \otimes R(0)=\left\vert 0\right\rangle_c\left\langle0\right\vert \otimes \sum_{n=0}^{\infty} \frac{n^n_m}{(1+n_m)^{n+1}}\left\vert n\right\rangle_m\left\langle n\right\vert \otimes R(0),
\end{eqnarray} 
where $R(0)$ is the reservoir state. The operation $U_D(t,0)$ and $U_K(t,0)$ keep the joint initial state $\chi(0)$ invariant. So the expectation value of a system operator $\hat{O}$ will be reduced to
\begin{eqnarray}
\mbox{Tr}_S\{\hat{O}\rho(t)\}&=&\mbox{Tr}_S\big\{\hat{O}~\mbox{Tr}_R\{U_S(t,0)\rho(0)\otimes R(0)U^\dagger_S(t,0)\}\big\}\nonumber\\
&=&\mbox{Tr}_{S\otimes R}\big\{U^\dagger_{OM}(t,0)U^\dagger_{E}(t,0)\hat{O}U_{E}(t,0)U_{OM}(t,0) \rho(0)\otimes R(0)\big\}. \label{fac}
\end{eqnarray} 

\section{Entanglement calculation}
We work in the regime of $g\ll 1$, so an initial Gaussian state will still be Gaussian after OMS evolution. For bipartite Gaussian states, the logarithmic negativity of the correlation matrix (CM) can well measure their entanglement. The CM is defined as
\begin{eqnarray}
\hat{V}= \left(
\begin{array}
[c]{cc}%
 \hat{A} & \hat{C} \\
 \hat{C}^T &  \hat{B}
\end{array}
\right).
\label{corr-matrix}
\end{eqnarray}
with the elements $\hat{V}_{ij}(t)=0.5\langle \hat{u}_i\hat{u}_j+\hat{u}_j\hat{u}_i\rangle-\langle \hat{u}_i\rangle\langle \hat{u}_j\rangle$, where $\hat{\vec{u}}=(\hat{x}_c(t),\hat{p}_c(t),\hat{x}_m(t),\hat{p}_m(t))^T$. The expectation values in the expression are calculated with Eq. (11). Then the corresponding logarithmic negativity is given as \cite{v-w, adesso, pl}
\begin{eqnarray}
E_{\cal N}=\mbox{max}[0, -\ln 2\eta^{-}],
\end{eqnarray}
where
\begin{eqnarray}
\eta^{-}=\frac{1}{\sqrt{2}}\sqrt{\Sigma-\sqrt{\Sigma^2-\mbox{det}\hat{V}}}
\end{eqnarray}
and
\begin{eqnarray}
\Sigma=\mbox{det}\hat{A}+\mbox{det}\hat{B}-2\mbox{det}\hat{C}.
\end{eqnarray}
To find the logarithmic negativity experimentally, one should measure the correlations $\langle \hat{u}_i\hat{u}_j\rangle$ for the cavity and mechanical modes. One proposal for doing so is discussed in \cite{ET-1}. More efforts should be taken to find the feasible ways to realize the measurement of logarithmic negativity. 

Since the drive operation $U_E(t,0)$ only displaces the system operators, it will not contribute to CM elements and can be neglected in the calculation of logarithmic negativity. Considering the expectation value in Eq. (\ref{fac}), what we should take into account is only the effective Hamiltonian in Eq. (\ref{eff}). Under this Hamiltonian the evolution of system operators are determined by the following differential equations:
\begin{eqnarray}
-i\frac{d\hat{a}}{d\tau }&=&ge^{-(\kappa +\gamma _{m})(t-\tau )/2}\sum_j D_j (\tau
)(\hat{b}e^{-i\omega _{m}\tau }+\hat{b}^{\dagger }e^{i\omega _{m}\tau }) \nonumber\\
&+&ge^{-\kappa (t-\tau )/2}\sum_j D_j (\tau )\big[\hat{n}_{m}(t,\tau )e^{i\omega _{m}\tau}+\hat{n}_{m}^{\dagger }(t,\tau )e^{-i\omega _{m}\tau}\big] \nonumber \\
-i\frac{d\hat{b}}{d\tau }&=&ge^{-(\kappa +\gamma _{m})(t-\tau )/2}e^{i\omega
_{m}\tau }\sum_j \big[D_j ^{\ast }(\tau )\hat{a}+D_j (\tau )\hat{a}^{\dagger }\big]\nonumber\\
&+&ge^{i\omega _{m}\tau -\frac{\gamma _{m}}{2}(t-\tau )}\sum_j \big[\hat{n}_{c}(t,\tau )D_j^{\ast }(\tau )+\hat{n}_{c}^{\dagger }(t,\tau )D_j(\tau )\big]
\nonumber \\
&+&ge^{i\omega _{m}\tau -\frac{\gamma _{m}}{2}(t-\tau )}\sum_j\sum_k D^{*}_j(\tau)D_k(\tau).
\label{om}
\end{eqnarray}%
The final term in the second equation can be neglected, since it only brings displacement which will not contribute to logarithmic negativity. The terms containing $\hat{a}$ and $\hat{b}$ on the right side of the equations indicate the beamsplitter (BS) effect between the cavity and mechanical modes, while those containing $\hat{a}^{\dagger}$ and $\hat{b}^{\dagger}$ indicate the squeezing effect which is the main cause of entanglement. Both two effects can be enhanced by the higher drive intensity, because each terms including the component $D(\tau)$. Moreover, one sees from the second term on the right side of each equation that a higher drive intensity will also enhance the noise effect. 

We rewrite Eq. (\ref{om}) in the form 
\begin{eqnarray}
\frac{d}{d\tau}\hat{\vec{u}}=\hat{M}(t,\tau)\hat{\vec{u}},
\label{VCM}
\end{eqnarray}
where
\begin{eqnarray}
\hat{M}(t,\tau)=\left(
\begin{array}
[c]{cccc}%
 0 & 0 & -\mathcal{I}\cos(\omega_m \tau) & -\mathcal{I}\sin(\omega_m \tau) \\
 0 &  0 & \mathcal{R}\cos(\omega_m \tau) & \mathcal{R}\sin(\omega_m \tau) \\
-\mathcal{R}\sin(\omega_m \tau) & -\mathcal{I}\sin(\omega_m \tau) & 0 &  0\\
\mathcal{R}\cos(\omega_m \tau) &\mathcal{I}\cos(\omega_m \tau) & 0 & 0
\end{array}
\right), \nonumber
\end{eqnarray}
where $\mathcal{R}=\Re e[P(t,\tau)]$ is the real part and $\mathcal{I}=\Im e[P(t,\tau)]$ is the imaginary part with $P(t,\tau)=2ge^{-(\kappa+\gamma_m)(t-\tau)/2} \sum_j D_j(\tau)$. Then the solution of above equation is found as
 \begin{eqnarray}
\hat{\vec{u}}(t)&=& e^{\int_0^t d\tau \hat{M}(t,\tau)}\hat{\vec{u}}(0) \nonumber\\
&=& \big(\cosh(\sqrt{m(t,0})\big)\hat{I}+\frac{\sinh\big(\sqrt{m(t,0)}\big)}{\sqrt{m(t,0)}}\hat{K}(t,0)\big)\hat{\vec{u}}(0),
\label{sol}
\end{eqnarray}
where $\hat{K}(t,0)=\int_0^t d\tau \hat{M}(t,\tau)$, and the function $m(t,0)$ is from the relation $\hat{K}^2(t,0)=m(t,0)\hat{I}$.

\section{Entanglement under different laser pulses}
\subsection{Entanglement evolution under a single laser pulse}

\begin{figure}[ptb]
\includegraphics[width=14.7cm]{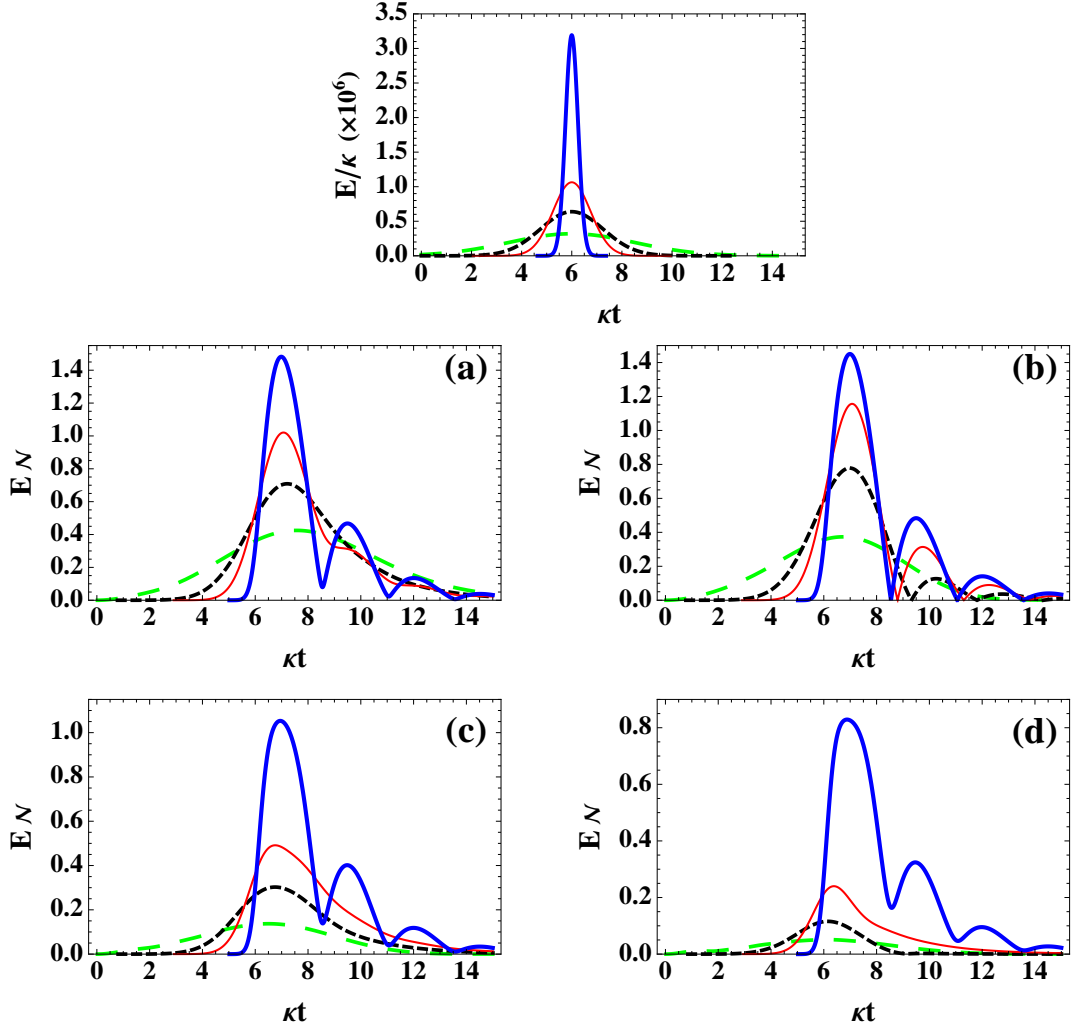}\caption{{Entanglement evolution under Gaussian pulse drive with different widths (displayed on the top). Long dashed (green) curve is for $\Delta \omega/\omega_m=0.4$, short dashed (black) curve for $\Delta \omega/\omega_m=0.8$, thin solid (red) curve for $\Delta \omega/\omega_m=4/3$, and thick solid (blue) curve for $\Delta \omega/\omega_m=4$, respectively. 
The corresponding detuning $\Delta_0/\omega_m$ for Figs.1(a)--1(d) are $\Delta_0/\omega_m=-1, -0.5, 0.5, 1$, respectively.
The other parameters are $g/\kappa=10^{-6}$, $\omega_m/\kappa=2.5$, $\omega_m/\gamma_m=10^7$, and $T=0$.}}%
\end{figure}

With the solution in Eq. (\ref{sol}), one can obtain the entanglement evolution under arbitrary pulsed drive. The first one we discuss is the Gaussian pulse, whose profile is $E(t-j_0)=\frac{E}{2\pi}e^{-\Delta\omega^2\cdot(t-j_0)^2/2}$. The pulses are displayed in the top of Fig. 1, with the different width $\Delta \omega=4\omega_m, 4/3 \omega_m, 0.8\omega_m, 0.4\omega_m$. 

One has the increased entanglement following the increase of pulse intensity. The entanglement keeps increasing though the pulse intensity has started to decrease. Only until the pulse almost disappears, does the entanglement begin to decrease and oscillate for a period. The entanglement will exist longer than the pulse duration. 

The entanglement under blue detuned central frequency [Fig. 1(a) and Fig.1(b)] is obviously larger than that under red detuned central frequency [Fig. 1(c) and Fig. 1(d)]. However, a main difference of the pulse driven entanglement from the CW counterpart is that a spectrum of frequencies with disparate detunings give rise to the actual entanglement. As seen in Fig. 1, the entanglement values for 
the pulses of narrower bandwidth behave monotonously in both blue and red detuned regime for the central frequency, while those of larger bandwidth show the entanglement patterns of both blue and red detuned frequency components. The entanglement from red detuned frequency components oscillates with time \cite{LH}, so this pattern always exists for the narrowest pulse (corresponding to the widest frequency bandwidth) in Fig. 1. As the result, the entanglement for the pulses of larger bandwidth show insignificant difference in Figs. 1(a) and 1(b), except that more oscillation pattern is introduced in 1(b) as the central frequency moves toward the red detuned regime. When the pulses' central frequency becomes red detuned, the oscillation for one of the pulses will disappear due to the suppression of entanglement for its whole frequency spectrum in the regime. On the other hand, given the same driving intensity, 
the entanglement will become higher with a narrower width. 

For comparison, we consider other pulses of different profiles, such as square pulse, triangle pulse, sawtooth pulse and trapezium pulse; see the top of Fig. 2. Their evolutions for the central frequency detuning $\Delta_0/\omega_m=-1$ [Fig. 2 (a)] are similar, so the different pulse shapes do not affect the entanglement much in this case. Meanwhile, difference appears in the evolution with red detuning $\Delta_0/\omega_m=1$ [Fig. 2 (b)]. Compared with Gaussian pulse driven OMS, the entanglement values under the pulse drives in Fig. 2 are higher around the SQ resonant point ($\Delta_0/\omega_m=-1$). It can be explained by the fact that a Gaussian pulse contains more significant off-peak frequency modes, and the frequency components whose detunings are far away from SQ resonant point contribute less to the entanglement. This is also the reason for why a narrower bandwidth of Gaussian pulse is better for entanglement creation. Moreover, due to their more centered frequency spectra, the differences of the entanglement evolutions in the blue and red detuned regime become more obvious for the pulses in Fig. 2.

\begin{figure}[ptb]
\includegraphics[width=14.7cm]{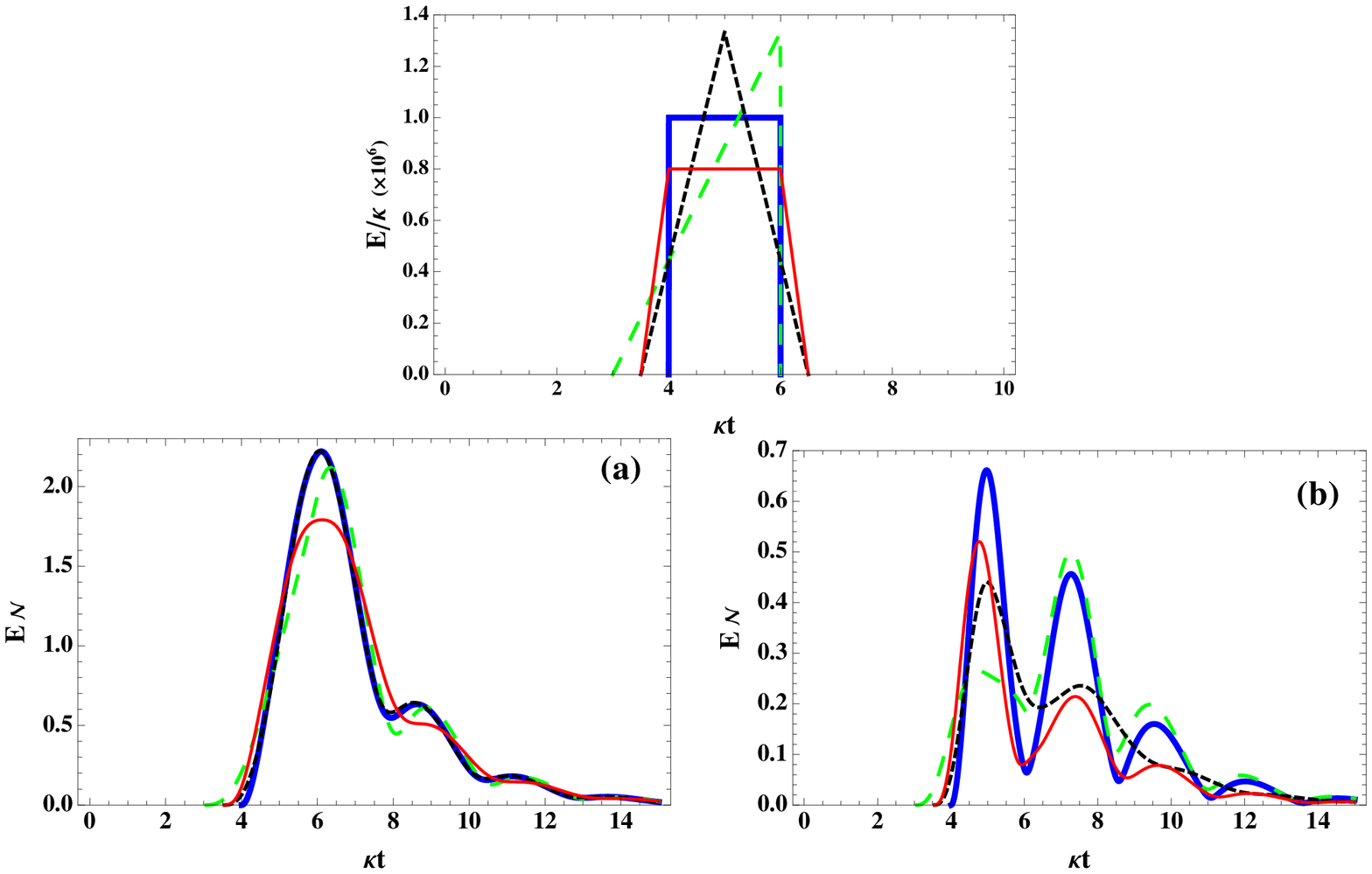}\caption{{Entanglement evolution under pulse drive with different profiles (displayed on the top). (a) The detuning is $\Delta_0/\omega_m=-1$. (b) The detuning is $\Delta_0/\omega_m=1$. The other parameters are the same as in Fig. 1.}}%
\end{figure}

\subsection{Entanglement evolution under a series of laser pulses}

As show in Fig. 1, the entanglement could last for longer time than the pulse duration. The entanglement may survive if one more pulse drive the OMS before it disappears. Since the entanglement will be kept and raised in this case, the minimal value of entanglement will be nonzero and the peak value of entanglement can be slight larger than that under a single pulse. This phenomenon will occur if the gap time between the centers of the pulses is not too large. In Fig. 3, we show the entanglement evolution of the OMS driven by a series of Gaussian pulses. The gap time ($\Delta (\kappa t)=8$) is relatively large in Fig. 3(a), resulting in the repeated evolution pattern of the entanglement under single pulse driven OMS [see Fig. 3(b)]. When the gap time ($\Delta (\kappa t)=5$) is relatively small as in Fig. 3(c), the entanglement evolution will be totally different as shown in Fig. 3(d). The peak value is about $1.6$ for the width $\Delta \omega/\omega_m=4$ and $1.1$ for the width $\Delta \omega/\omega_m=4/3$, which are slightly larger than the ones ($1.5$ and $1$, respectively) shown in Fig. 1. At the same time, the minimal value becomes nonzero (about $0.1$) for the width $\Delta \omega/\omega_m=4$. Especially, for the width $\Delta \omega/\omega_m=4/3$ (the overlap of two pulses can be neglected in this case), the minimal value will be even larger (about $0.3$). A rather stable entanglement ($>0.3$) can therefore be achieved by pulsed drive, even the overlap of two pulses is not so obvious. 

\begin{figure}[ptb]
\includegraphics[width=14.7cm]{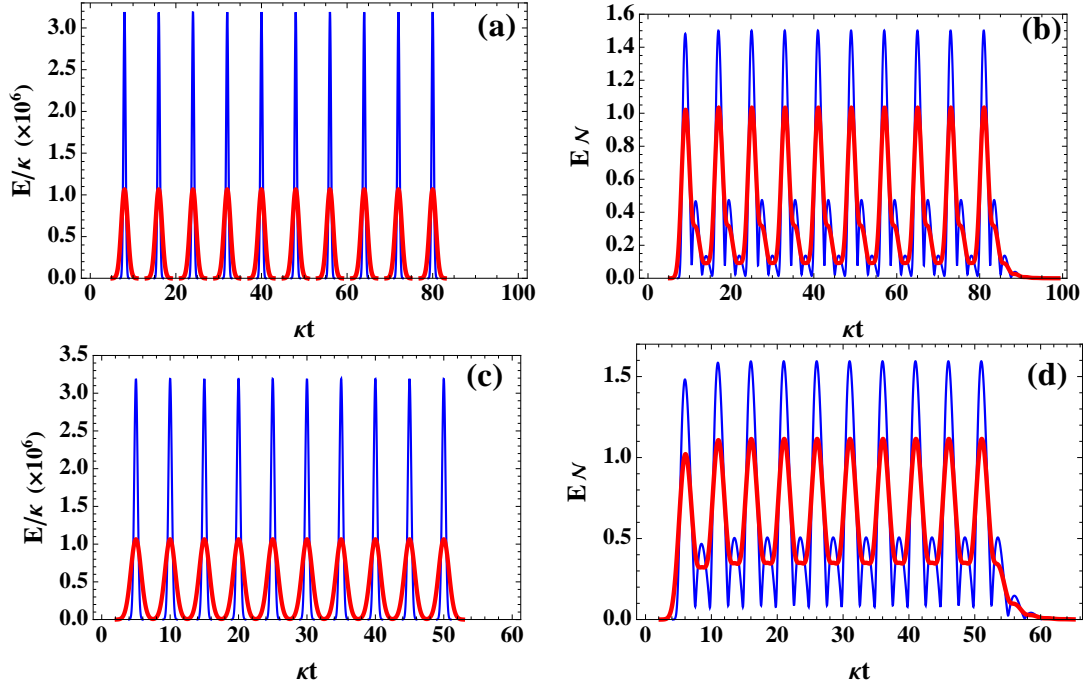}\caption{{Entanglement evolution under a series of Gaussian pulses with the detuning $\Delta_0/\omega_m=-1$. The left two figures are for the pulses and the right ones are for the entanglement evolution. The gap time of the pulse centers in (a) and (b) are $\Delta (\kappa t)=8$ and $\Delta (\kappa t)=5$, respectively. The width for thin (blue) curve is $\Delta \omega/\omega_m=4$, while that for thick (red) curve is $\Delta \omega/\omega_m=4/3$. The other parameters are the same as in Fig. 1.}}%
\end{figure}

\subsection{Entanglement evolution under high temperature}

\begin{figure}[ptb]
\includegraphics[width=14.7cm]{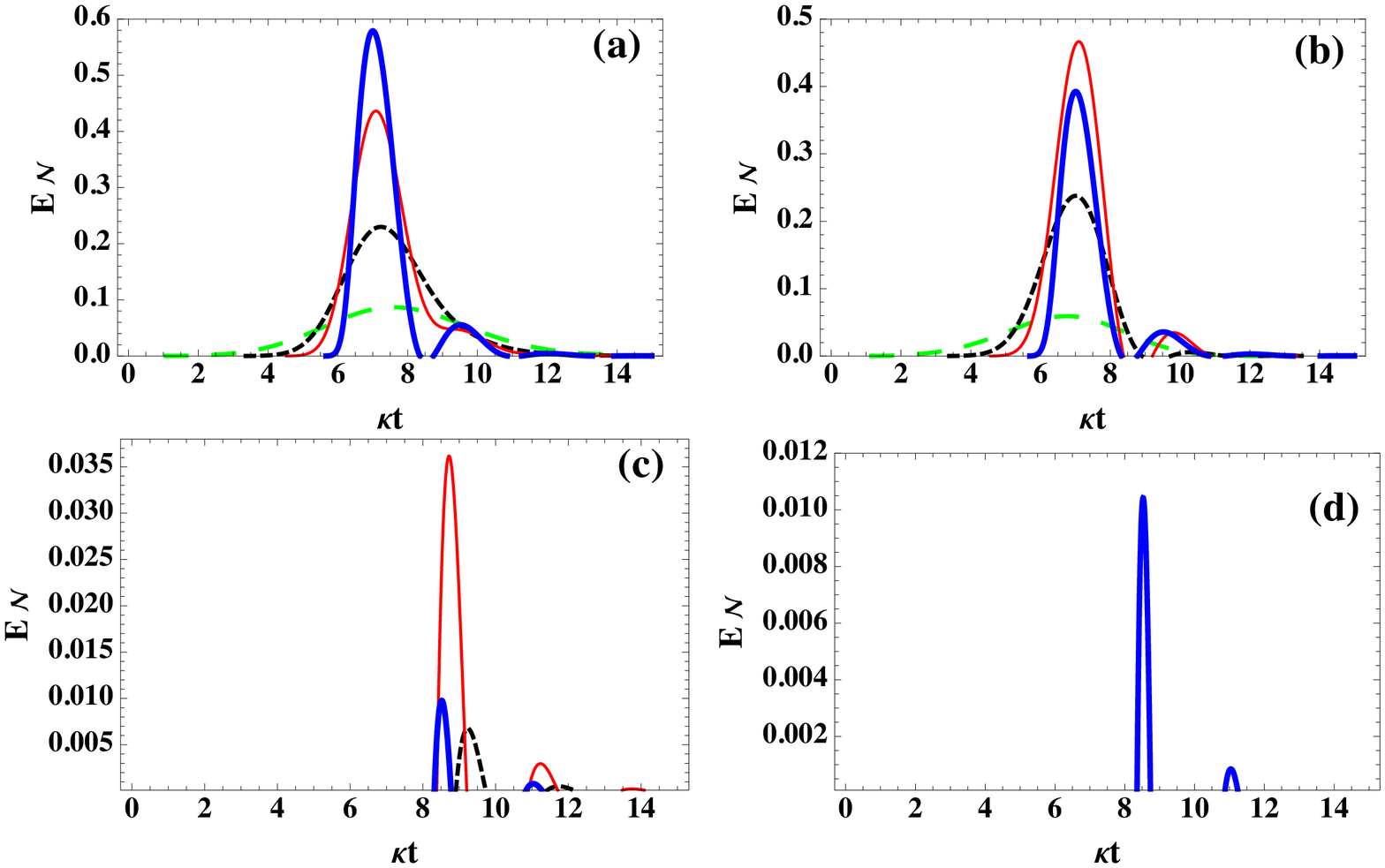}\caption{{Entanglement evolution under Gaussian pulse drive with the temperature $n_m=10^4$. The other parameters are the same to Fig.1.}}%
\end{figure}

In the above discussions, the temperature is set to be $T=0$. However, temperature is also a main factor that affects the existence of entanglement. In Fig. 4 we show the entanglement evolution under the condition $n_m=10^4$ while other conditions are the same as in Fig. 1. Obviously, the entanglement decreases with high temperature, and the phenomenon of entanglement sudden death and sudden 
revival becomes apparent. Through the comparison, one sees that the robustness of the entanglement against high temperature for the blue detuned frequency components is much more significant than that from red detuned ones. As it is also shown in these figures, the entanglement evolutions under higher temperature become more complicated, because they result from more intricate interplay of various physical factors (temperature, frequency bandwidth, central frequency detuning and drive intensity). The competition between the factors gives rise to the seemingly unexpected entanglement amplitudes in Figs. 4(b), 4(c) and 4(d). At the central frequency detunings in 4(b) and 4(c), the higher driving intensity of the pulse with the widest bandwidth actually suppresses the entanglement value, because the thermal noise effect killing the entanglement is enhanced by the driving strength [see Eq. (15)]. If the central frequency is exactly BS resonant as in 4(d), the suppression of entanglement for the whole spectrum becomes more significant with the temperature, leaving the remnant due to the far-away frequency components from the point, which is only possible for the pulse with the widest bandwidth.  
To achieve high and robust entanglement, blue detuning around $\Delta_0/\omega_m=-1$ associated with narrow pulse bandwidth, should be 
the best choice.

\section{Conclusion}

We have depicted the optomechanical entanglement evolutions under various pulsed drives. Compared with CW drives, high and robust entanglement can also be created especially with pulses of blue detuning and narrow bandwidth. Such entanglement can last for long time with repeated series of driving pulses. The flexibility in engineering pulse frequencies and shapes enables one to achieve variety of entanglement evolution patterns for OMSs.

\section*{Acknowledgments}
Q. L. was funded by National Natural Science Foundation of China (11005040), Natural Science Foundation of Fujian Province of China (2014J01015), Program for
New Century Excellent Talents in Fujian Province University (2012FJ-NCET-ZR04) and Promotion Program for Young and Middle-aged Teacher in
Science and Technology Research of Huaqiao University (ZQN-PY113).

\end{document}